\documentclass[a4paper,11pt]{article}
\usepackage[margin=25mm]{geometry}
\usepackage{fontspec}
\usepackage{amsmath,amssymb,graphicx}
\usepackage[font=small]{caption}
\usepackage{tikz}
\usetikzlibrary{arrows.meta,decorations.pathreplacing}
\newcommand{\zRR}{z_{\mathrm{RR}}}
\newcommand{\SA}{\mathsf{SA}}
\newcommand{\ISA}{\mathsf{ISA}}

\newcommand{\LCP}{\mathsf{LCP}}
\newcommand{\lcp}{\operatorname{lcp}}
\newcommand{\LTP}{\operatorname{LTP}}

\newcommand{\LZ}{\operatorname{LZ}}

\newcommand{\Occ}{\operatorname{Occ}}
\newcommand{\dist}{\operatorname{dist}}

\newcommand{\LSB}{\mathsf{LSB}}
\newcommand{\Root}{\operatorname{root}}

\newcommand{\BIN}{\operatorname{BIN}}

\newcommand{\NUM}{\operatorname{NUM}}

\newcommand{\Dep}{\leadsto}

\newcounter{fibproperty}
\renewcommand{\thefibproperty}{\roman{fibproperty}}

\input{theorems.tex}
\usepackage[hidelinks,pdfversion=1.7]{hyperref}
\title{Tight Logarithmic Separations between LZ77 and LZRR}
\author{Yuto Fujie\thanks{\href{mailto:fujie.yuto.104@s.kyushu-u.ac.jp}{\texttt{fujie.yuto.104@s.kyushu-u.ac.jp}}}\\[3pt]
  \small Joint Graduate School of Mathematics for Innovation, Kyushu University, Japan}
\date{\today}
\begin{document}
\maketitle
\begin{abstract}
LZRR is a copy-based lossless compression method that produces a bidirectional macro scheme. It
greedily parses a string from left to right. At each step, it can copy from an occurrence on either
side of the current position, provided that the copy creates no cycle. Let $z$ and $\zRR$ denote the
sizes of the LZ77 and LZRR parsings, respectively. Under the original LZRR rule, we show that
$z/\zRR=\Theta(\log n)$ on a family of Fibonacci words and that $\zRR/z=\Theta(\log n)$ on a family
based on bit reversal, where $n$ is the string length. The Fibonacci separation holds under every
tie-breaking rule. Both separations are asymptotically tight.
\end{abstract}

\section{Introduction}
\label{sec:intro}

Lossless data compression has long been a fundamental tool for storing and transmitting large
amounts of data. LZ77, introduced by Ziv and Lempel in 1977, is one of the classical and most widely
studied lossless compression methods~\cite{ziv1977universal}. It parses a string greedily from left
to right, copying the longest match from an earlier occurrence whenever possible. More recently, the
compression of highly repetitive collections, such as genomic databases and histories of document or
software versions, has received considerable attention. Methods that exploit repeated substrings are
particularly effective for such data, and LZ77 plays a central
role~\cite{kociumaka2023repetitive,navarro2021ordered}. We write $z(T)$ for the size of the LZ77
parsing of a string $T$, measured in phrases. This size is also widely used as a measure of
repetitiveness.

A bidirectional macro scheme (BMS) generalizes LZ77 by allowing copies from occurrences on either
side, provided that the references are acyclic and the original string can be
recovered~\cite{storer1982substitution}. Let $b(T)$ be the size of a smallest BMS for $T$. Since
every LZ77 parsing is a valid BMS, $b(T)\le z(T)$, and the upper bound $z(T)=O(b(T)\log |T|)$ is
asymptotically tight~\cite{navarro2021ordered}. Finding a smallest BMS is
NP-hard~\cite{storer1982substitution,navarro2021ordered}, motivating efficient methods for
constructing small bidirectional parsings. Examples include the lex-parse~\cite{navarro2021ordered},
lcpcomp~\cite{dinklage2017tudocomp}, and BMSs derived from BWT runs~\cite{navarro2021ordered}. In
this paper, we focus on LZRR, introduced by Nishimoto and Tabei~\cite{nishimoto2022lzrr}.

LZRR retains the greedy left-to-right parsing strategy of LZ77, but allows the source of a phrase to
occur on either side of the current position. At each step, it chooses a longest copy that preserves
the validity of the current bidirectional parse. We write $\zRR^\tau(T)$ for the size of the
resulting parsing when ties between longest valid copies are resolved according to a rule $\tau$,
and $\zRR(T)$ for its size under the candidate order and source-update rule of Nishimoto and Tabei
(Definition~\ref{def:lzrr-candidate-order}).

Nishimoto and Tabei proved $\zRR(T)\le z(T^R)$, where $T^R$ is the reversal of
$T$~\cite[Theorem~3]{nishimoto2022lzrr}. Applying this inequality in both orientations gives
\[
\min\{\zRR(T),\zRR(T^R)\}
\le\min\{z(T),z(T^R)\}.
\]
Their experiments use this convention, running each method on both orientations and reporting the
smaller parsing size~\cite[Section~5]{nishimoto2022lzrr}. The proof of the upper bound applies to
every choice among longest valid copies, so $\zRR^\tau(T)\le z(T^R)$ holds for every $\tau$.
Together with $z(T)=O(b(T)\log |T|)$ and $b(T^R)=b(T)$, this also gives $\zRR^\tau(T)=O(b(T)\log
|T|)$~\cite{navarro2021ordered,bannai2026reversal}. Thus both methods approximate the smallest BMS
size within a logarithmic factor.

To determine whether allowing right references gives LZRR an asymptotic advantage over LZ77, we must
compare the two methods with the input orientation fixed. Reversing a string can itself reduce its
LZ77 parsing size by a logarithmic factor~\cite{shibata2026lz}. Thus, the guarantee $\zRR(T)\le
z(T^R)$ does not establish that the additional source choices of LZRR improve compression relative
to $z(T)$. We therefore compare $\zRR(T)$ directly with $z(T)$ to assess the effect of allowing
right references. Although LZRR considers occurrences on both sides, its earlier source choices can
create dependencies that make later copies invalid. We show that this flexibility can yield a
logarithmic improvement over LZ77, but can also lead to a parsing that is logarithmically larger.

The following string families give the separations in both directions.
\begin{itemize}
\item \textbf{LZRR can be asymptotically smaller than LZ77.} For every Fibonacci word $F_k$ with
integer $k\ge8$ and every $\tau$, we prove $\zRR^\tau(F_k)=5$
(Theorem~\ref{thm:fib-forward-constant}). The known bound $z(F_k)=\Theta(\log |F_k|)$ gives
$z(F_k)/\zRR^\tau(F_k)=\Theta(\log |F_k|)$.
\item \textbf{LZ77 can be asymptotically smaller than LZRR.} We construct a family $T_h$ based on
bit reversal for which $\zRR(T_h)/z(T_h)=\Theta(\log |T_h|)$ under the original tie-breaking rule
(Theorem~\ref{thm:bitrev-separation}). Its alphabet size is $\Theta(\sqrt{|T_h|})$.
\end{itemize}
Lemma~\ref{lem:z-zrr-general-ratios} bounds both ratios by $O(\log |T|)$ for every $\tau$. Both
separations are therefore asymptotically tight. In particular, under the original LZRR rule, neither
$z(T)$ nor $\zRR(T)$ is bounded by a constant multiple of the other.

Our analysis of Fibonacci words uses known LZ77 bounds and descriptions of substring
occurrences~\cite{navarro2021ordered,bannai2026reversal,guo2025net}. The original implementation
also reports five LZRR phrases on the benchmark \texttt{fib41}~\cite{nishimotoLzrrCode}. For the
opposite separation, we adapt a block construction based on bit
reversal~\cite{shibata2026lz,bille2026endvs77} and analyze the candidate order and the resulting
dependencies.

\section{Preliminaries}
\label{sec:preliminaries}

\subsection{Strings}
Let $\Sigma$ be a finite, totally ordered alphabet. For a string $T\in\Sigma^*$ of length $n=|T|$,
positions range from $0$ to $n-1$. We write $\varepsilon$ for the empty string and $UV$ for
concatenation. The substring $T[i..j]$ includes both endpoints, and $T[i..i-1]=\varepsilon$. We
abbreviate the suffix $T[i..n-1]$ to $T[i..]$ and write $T^R$ for the reversal. The occurrence set
is $\Occ(U,T)=\{p:0\le p\le n-|U|,\ T[p..p+|U|-1]=U\}$. We write $\prec$ for lexicographic order, in
which a proper prefix precedes the full string. All logarithms have base two.

For $d\ge0$ and $0\le t<2^d$, let $\BIN_d(t)$ be the $d$-bit representation of $t$, with leading
zeros, and let $\NUM(w)$ be the integer represented by the binary string $w$. Set
$\BIN_0(0)=\varepsilon$ and $\NUM(\varepsilon)=0$.

\subsection{Suffix arrays and longest common prefixes}
The suffix array $\SA$ is the permutation of text positions satisfying $T[\SA[0]..]\prec\cdots\prec
T[\SA[n-1]..]$~\cite{manber1993suffix}. Its inverse $\ISA$ satisfies $\ISA[\SA[r]]=r$. Write
$\lcp(U,V)$ for the length of the longest common prefix of strings $U,V$, and set
$\lcp_T(p,q)=\lcp(T[p..],T[q..])$. We omit the subscript when $T$ is clear. The LCP array is given
by $\LCP[0]=0$ and $\LCP[r]=\lcp_T(\SA[r-1],\SA[r])$ for $1\le r<n$.

\subsection{Bidirectional macro schemes}
Let $T=B_1\cdots B_s$ be a parsing of $T$ into nonempty phrases. Write $p_i$ and $\ell_i$ for the
start and length of $B_i$. An explicit phrase has $\ell_i=1$ and stores $T[p_i]$ directly. A copy
phrase specifies a source position $q_i\ne p_i$ such that
$T[p_i..p_i+\ell_i-1]=T[q_i..q_i+\ell_i-1]$. The source and target may overlap, and $q_i$ can be any
position where the source occurs.

\begin{definition}[Dependency graph]\label{def:dependency-graph}
For a parsing $\mathcal P$ with specified sources, its dependency graph $G_{\mathcal P}$ has the
text positions as vertices. Each copy phrase contributes an edge $p_i+j\to q_i+j$ for each $0\le
j<\ell_i$. Explicit phrase positions have no outgoing edges.
\end{definition}
A parsing is a valid bidirectional macro scheme (BMS) if its dependency graph is
acyclic~\cite{storer1982substitution,navarro2021ordered,nishimoto2022lzrr}. Its size is the number
$s$ of phrases. We write $b(T)$ for the size of the smallest BMS for $T$.

In a valid parsing, every weakly connected component has one root with no outgoing edge. Following
references from any position in that component leads to its root. We use component to mean weakly
connected component. Write $\Root_{\mathcal P}(u)$ for the root reached from $u$, and write
$u\sim_{\mathcal P}v$ when $u$ and $v$ reach the same root. Write $u\Dep_{\mathcal P}v$ if there is
a directed path of length zero or more from $u$ to $v$. We omit the subscripts when the parsing is
clear. Adding an edge from a root $u$ to a position $v$ in its component closes a cycle along the
existing path $v\Dep u$.

\paragraph{Example.}
Let $T=\mathtt{aabaabaaa}$. Consider the parsing
$(\mathtt{aabaa},\mathtt{b},\mathtt{aa},\mathtt{a})$. The phrase at position 0 copies from position
3, and the phrase at position 6 copies from position 7. Positions 5 and 8 are explicit. For example,
the paths $0\to3\to6\to7\to8$ and $2\to5$ reach these explicit characters. Every other position also
reaches one of them, so the BMS is valid and $b(T)\le4$.

\subsection{LZ77}
LZ77 greedily parses the string from left to right~\cite{ziv1977universal,navarro2021ordered}. At
the current position $p$, it chooses a source $q<p$ maximizing $\lcp_T(p,q)$ and copies that many
characters. If there is no positive-length copy, it stores $T[p]$ explicitly. The source and target
may overlap. We write $\LZ(T)$ for the resulting parsing and $z(T)=|\LZ(T)|$ for its size. Ties
between sources leave the phrase boundaries unchanged. LZ77 achieves the smallest size among
parsings whose copy sources start to the left of their targets~\cite{navarro2021ordered}.

\paragraph{Example.}
For $T=\mathtt{aabaabaaa}$, the LZ77 parsing is
$\LZ(T)=(\mathtt{a},\mathtt{a},\mathtt{b},\mathtt{aabaa},\mathtt{a})$. Positions 0 and 2 are
explicit. The copy phrases at positions 1, 3, and 8 can all use position 0 as their source. At
position 3, the copy overlaps its source and matches five characters before the next mismatch. Thus
$z(T)=5$.

\subsection{LZRR}
LZRR processes phrases from left to right~\cite{nishimoto2022lzrr}. At the current position $p$, let
$\mathcal P$ contain the processed phrases before $p$ and explicit one-character phrases at the
unprocessed positions from $p$ onward. For $q\ne p$, let $\mathcal P_t$ be the parsing obtained by
copying the $t$ characters starting at $p$ from $q$, and set $\mathcal P_0=\mathcal P$. The length
of the longest valid copy from $q$ is
\[
\LTP(\mathcal P,p,q)=\max\{t:0\le t\le\lcp_T(p,q),\ \mathcal P_t\text{ is valid}\}.
\]
The algorithm chooses a source maximizing $\LTP(\mathcal P,p,q)$ if the maximum is positive.
Otherwise it stores $T[p]$ explicitly.

A tie-breaking rule $\tau$ chooses one of the sources giving a longest valid copy, and may use the
current parsing to make that choice. The choice can affect later phrases. Write $\zRR^\tau(T)$ for
the size of the LZRR parsing obtained under $\tau$. We follow the candidate order and source-update
rule used in the journal version of Nishimoto and Tabei~\cite{nishimoto2022lzrr}. The following
definition restates their rule in our notation. For the current position $p$ and candidate $q$,
define
\[
\dist_p(q)=\begin{cases}
\ISA[q]-\ISA[p],&\ISA[q]\ge\ISA[p],\\
\ISA[p]-\ISA[q]-\tfrac12,&\ISA[q]<\ISA[p].
\end{cases}
\]
\begin{definition}[Nishimoto--Tabei source selection]\label{def:lzrr-candidate-order}
Examine candidates $q\ne p$ in nonincreasing order of $\lcp_T(p,q)$, breaking ties by increasing
$\dist_p(q)$. Update the chosen source only when $\LTP(\mathcal P,p,q)$ strictly improves the best
length found so far. Stop when the next candidate's common prefix length is at most the best length.
\end{definition}
Among candidates with equal common prefix lengths, the rule gives priority to a smaller suffix-array
distance and then to the lexicographically smaller side. Since an equal valid copy length does not
trigger an update, it selects the first candidate in this order that gives a longest valid copy.
Thus $\zRR(T)$ denotes the parsing size under this Nishimoto--Tabei rule. The result in
Section~\ref{sec:fib} holds for every $\tau$, and the lower bound in Section~\ref{sec:bit_rev} uses
this original rule.

\paragraph{Example.}
For $T=\mathtt{aabaabaaa}$, LZRR first copies $\mathtt{aabaa}$ from position 3. The only other
$\mathtt{b}$ is at position 2, so copying it at position 5 would create the cycle $2\to5\to2$.
Position 5 is therefore explicit. At position 6, copying $\mathtt{aa}$ from position 3 would create
the cycle $3\to6\to3$. The copy from position 7 is valid and overlaps the target. Finally, every
other $\mathtt{a}$ already leads to position 8, so position 8 is also explicit. This gives the same
four-phrase parsing as in the BMS example.

\paragraph{Dependencies during parsing.}
Let $\mathcal P$ contain the processed phrases before the current position $p$ and explicit
one-character phrases at all positions from $p$ onward. Each position at or after $p$ is then the
root of a component, and distinct such positions belong to distinct components. This remains true
even if earlier processed phrases refer to these positions. Processing a phrase only adds copy edges
and never deletes existing edges. Hence two positions in the same component can never become
separated later.

\begin{lemma}[Copy length from a position to the right]\label{lem:future-copy}
For the parsing $\mathcal P$ above and every $q>p$, we have $\LTP(\mathcal P,p,q)=\lcp_T(p,q)$.
\end{lemma}
\begin{proof}
Let $t\le\lcp_T(p,q)$. A length-$t$ copy adds the edges $p+j\to q+j$ for $0\le j<t$. Contract each
existing component to a single vertex. Each added edge then increases the root position by $q-p>0$.
Thus a path containing new edges cannot return to its original component, so no directed cycle is
created. The entire common prefix can therefore be copied. Conversely, no copy can exceed the common
prefix length.
\end{proof}

\begin{lemma}\label{lem:unique-sink-tree}
Let $H$ be a finite directed acyclic graph in which a vertex $r$ has outdegree zero and every other
vertex has outdegree one. Then following outgoing edges from any vertex reaches $r$. Thus, adding an
edge $r\to u$ for any vertex $u$ creates a directed cycle.
\end{lemma}
\begin{proof}
Since $H$ is finite and acyclic, following outgoing edges must end at a vertex of outdegree zero,
which can only be $r$. Adding $r\to u$ closes a directed cycle along the existing path from $u$ to
$r$ (a self-loop if $u=r$).
\end{proof}

\subsection{General upper bounds}
Both LZ77 and LZRR under any rule $\tau$ give valid BMSs, so $b(T)\le z(T)$ and
$b(T)\le\zRR^\tau(T)$. Reversing the phrases and references of a BMS preserves its validity and
size, giving $b(T^R)=b(T)$~\cite{bannai2026reversal}.
\begin{lemma}\label{lem:z-zrr-general-ratios}
For every string $T$ and every tie-breaking rule $\tau$, we have $\zRR^\tau(T)\le z(T^R)$. If
$n=|T|\ge2$, then $z(T)=O(\zRR^\tau(T)\log n)$ and $\zRR^\tau(T)=O(z(T)\log n)$.
\end{lemma}
\begin{proof}
The argument of Nishimoto and Tabei~\cite[Theorem~3]{nishimoto2022lzrr} applies to any choice among
longest valid copies. Reverse the LZ77 parsing of $T^R$ to obtain a parsing of $T$ into $z(T^R)$
phrases with only rightward references. Let $p$ be the start of an LZRR phrase, and consider the
phrase of the reversed parsing that contains $p$. If it is a copy phrase, its suffix starting at $p$
still has a source to the right. By Lemma~\ref{lem:future-copy}, this copy is valid, so the greedy
LZRR phrase reaches at least the end of that phrase. If it is explicit, advancing by one character
suffices. Thus each phrase of the reversed parsing contains at most one LZRR phrase start, giving
$\zRR^\tau(T)\le z(T^R)$.

The known bound $z(T)=O(b(T)\log n)$~\cite{navarro2021ordered} and $b(T)\le\zRR^\tau(T)$ give the
first ratio bound. The second follows from $\zRR^\tau(T)\le z(T^R)=O(b(T^R)\log n)$ and
$b(T^R)=b(T)\le z(T)$.
\end{proof}

\section{When LZRR Is Smaller than LZ77}
\label{sec:fib}

We use Fibonacci words to obtain a logarithmic separation with $\zRR\ll z$. Define the Fibonacci
words by $F_1=b$, $F_2=a$, and $F_j=F_{j-1}F_{j-2}$ for $j\ge 3$. Write $f_j=|F_j|$ and
$C_j=F_j[0..f_j-3]$ for $j\ge 3$. We call $C_j$ the central word, with $C_3=\varepsilon$. Assume
$a\prec b$ and let $k\ge 8$. Positions are numbered from zero. Write $\mathtt{uv}$ for the last two
characters of $F_k$. Here $\mathtt{u}$ and $\mathtt{v}$ each denote a single character, with the
assignment of $a$ and $b$ depending on the parity of $k$: $(\mathtt{u},\mathtt{v})=(a,b)$ for odd
$k$ and $(\mathtt{u},\mathtt{v})=(b,a)$ for even $k$.

We collect the known properties needed below. The recurrence implies that $F_j$ is a prefix of
$F_{j+1}$ for $j\ge2$.
\begin{lemma}\label{lem:fib-known-properties}
\leavevmode\par
\begin{list}{\textup{(\thefibproperty)}}{%
\usecounter{fibproperty}%
\setlength{\leftmargin}{1.9em}%
\setlength{\labelwidth}{1.5em}%
\setlength{\labelsep}{0.4em}%
\setlength{\topsep}{2pt}%
\setlength{\partopsep}{0pt}%
\setlength{\itemsep}{0pt}%
\setlength{\parsep}{0pt}}
\item\label{item:fib-coprime} If $j\ge3$, then $\gcd(f_j,f_{j-1})=1$.
\item\label{item:fib-occ-two} If $j\ge6$, then $\Occ(F_{j-2},F_j)=\{0,\allowbreak
f_{j-2},\allowbreak f_{j-1}\}$.
\item\label{item:fib-occ-four} If $j\ge8$, then $\Occ(F_{j-4},F_j)=\{0,\allowbreak
f_{j-4},\allowbreak f_{j-3},\allowbreak f_{j-2},\allowbreak f_{j-2}+f_{j-4},\allowbreak
f_{j-1},\allowbreak f_{j-1}+f_{j-4},\allowbreak f_j-f_{j-4}\}$.
\item\label{item:fib-commutation} For $j\ge3$, \mbox{\(\displaystyle
F_j=C_j\cdot
\begin{cases}
ab,&\text{if }j\text{ is odd},\\
ba,&\text{if }j\text{ is even};
\end{cases}
\qquad
F_{j-2}F_{j-1}=C_j\cdot
\begin{cases}
ba,&\text{if }j\text{ is odd},\\
ab,&\text{if }j\text{ is even}.
\end{cases}\)}
\end{list}
\end{lemma}
Items (ii) and (iii) follow from Theorem~17 of Guo and Kishi~\cite{guo2025net} with index
differences of two and four, respectively. Occurrence positions have been converted from one-based
to zero-based indexing. Item (i) follows by repeatedly applying
$\gcd(f_j,f_{j-1})=\gcd(f_{j-1},f_{j-2})$. For completeness, item (iv) follows by induction from the
recurrence. The case $j=3$ is $F_3=ab$ and $F_1F_2=ba$. For $j\ge4$, apply the induction hypothesis
to the parenthesized factors in $F_j=F_{j-2}(F_{j-3}F_{j-2})$ and
$F_{j-2}F_{j-1}=F_{j-2}(F_{j-2}F_{j-3})$. The two strings agree except in their last two characters,
whose order follows from the parity of $j$.

We also use the following elementary property of residue classes.
\begin{lemma}\label{lem:fib-modular-shift}
Let $A,B$ be positive integers with $\gcd(A,B)=1$, and let $d\in\mathbb Z$. On the integer interval
$J=[d,d+A)$, define $\phi(x)=d+((x-d-B)\bmod A)$. For every $x\in J$, the sequence
$x,\phi(x),\phi^2(x),\ldots$ visits every position of $J$ exactly once before returning to $x$.
\end{lemma}
\begin{proof}
For every $t\ge0$, we have $\phi^t(x)\equiv x-tB\pmod A$. If $\phi^r(x)=\phi^t(x)$, then
$(r-t)B\equiv0\pmod A$. Since $\gcd(A,B)=1$, this implies $r\equiv t\pmod A$, and hence $r=t$ when
$0\le r,t<A$. Thus $x,\phi(x),\ldots,\phi^{A-1}(x)$ visit all $A$ positions of $J$. Moreover,
$\phi^A(x)$ and $x$ both belong to $J$ and are congruent modulo $A$, so they are equal.
\end{proof}

\begin{theorem}\label{thm:fib-forward-constant}
For every integer $k\ge8$ and every tie-breaking rule $\tau$, $\zRR^\tau(F_k)=5$. In particular,
$\zRR(F_k)=5$.
\end{theorem}

\begin{proof}

We show that the strings represented by the LZRR phrases are, in order,
\[
(C_{k-1},\,\mathtt{vu}F_{k-4},\,C_{k-3},\,\mathtt{u},\,\mathtt{v}).
\]
Figure~\ref{fig:fib-source-overview} illustrates the relative positions of the first three phrases
and a source for each of them on $F_9$. The sources of the first two phrases start to the right of
their targets, whereas the source used below for the third phrase starts to the left of its target.
We examine the three copy phrases in turn.

\begin{figure}[!tbp]
\centering
\includegraphics[width=.95\linewidth]{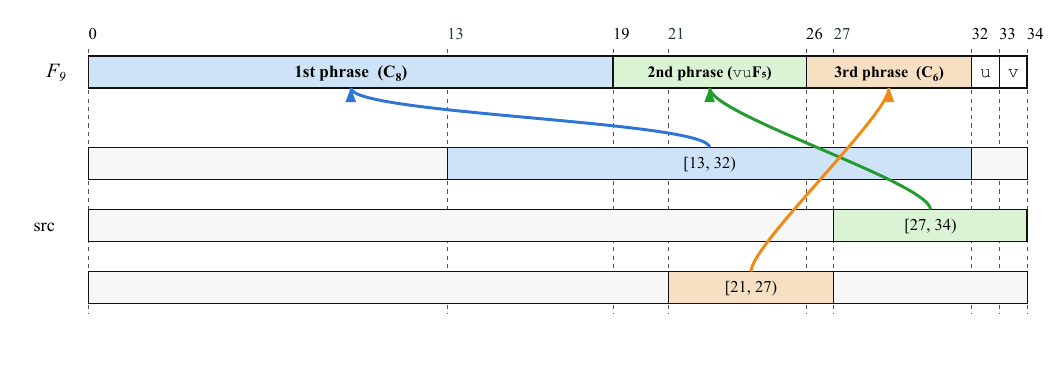}
\caption{The first three phrases and their sources on $F_9$. Positions are numbered from zero and
intervals are half-open. All rows use the same coordinate scale. Arrows indicate copying from a
source interval to its target. The third source is one valid choice.}
\label{fig:fib-source-overview}
\end{figure}

For the first phrase, Lemma~\ref{lem:fib-known-properties}(\ref{item:fib-commutation}) gives
$\lcp_{F_k}(0,f_{k-2})=f_{k-1}-2.$
The source at $f_{k-2}$ starts to the right of the target, so a copy of this length is valid by
Lemma~\ref{lem:future-copy}. Since $f_{k-1}-2>f_{k-2}$, any other source attaining this length must
also start with $F_{k-2}$. By Lemma~\ref{lem:fib-known-properties}(\ref{item:fib-occ-two}),
$F_{k-2}$ occurs in $F_k$ only at positions $0$, $f_{k-2}$, and $f_{k-1}$. Position $0$ is the
target itself, and $f_{k-2}$ is the source just given. The suffix starting at $f_{k-1}$ has length
only $f_{k-2}$ and is therefore too short. Thus the first phrase has the unique source $f_{k-2}$ and
length $f_{k-1}-2$.

\begin{figure}[!tbp]
\centering
\includegraphics[width=.85\linewidth,trim=10bp 10bp 15bp 5bp,clip]{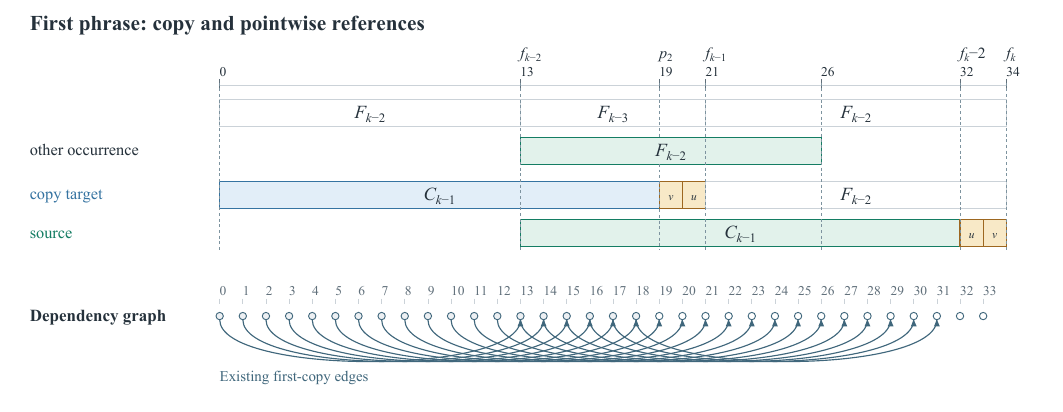}
\caption{The first phrase on $F_9$, with target starting at 0 and unique source starting at
$13=f_7$. Each dependency edge points from a copied position to its source.}
\label{fig:fib-first-copy}
\end{figure}

The second phrase starts at position $f_{k-1}-2$, where
$F_k[f_{k-1}-2\,..\,f_{k-1}+f_{k-4}-1]=\mathtt{vu}F_{k-4}.$ The occurrences of $F_{k-4}$ are given
by Lemma~\ref{lem:fib-known-properties}(\ref{item:fib-occ-four}). Among these occurrences, the two
preceding characters are $\mathtt{vu}$ exactly when the occurrence of $F_{k-4}$ starts at
$f_{k-3},f_{k-1},f_k-f_{k-4}.$ Hence, besides the target itself, the two possible sources of
$\mathtt{vu}F_{k-4}$ start at $f_{k-3}-2$ and $f_k-f_{k-4}-2$. The first phrase has already
introduced the reference $f_{k-3}-2\longrightarrow f_{k-1}-2.$ Choosing the left source would
introduce the reverse reference and hence create a directed cycle of length two. In contrast, the
source at $f_k-f_{k-4}-2$ starts to the right of the target and therefore gives a valid copy by
Lemma~\ref{lem:future-copy}. This copy has length $f_{k-4}+2$ and reaches the end of the string, so
it cannot be extended further. Thus the second phrase has the unique source $f_k-f_{k-4}-2$ and
length $f_{k-4}+2$.

\begin{figure}[!tbp]
\centering
\includegraphics[width=.95\linewidth]{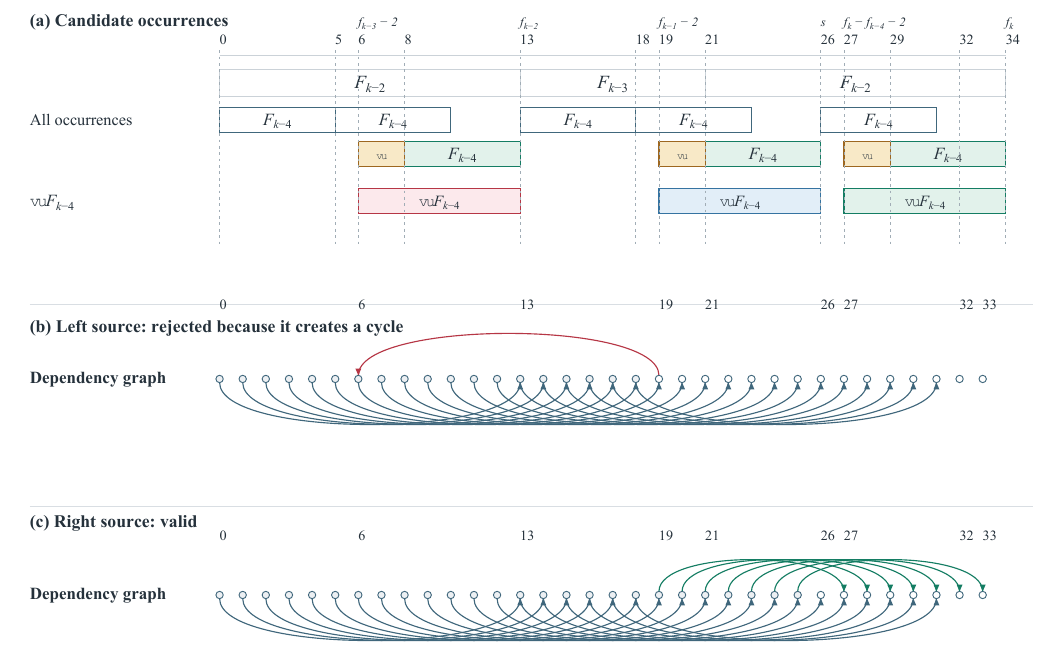}
\caption{The second phrase on $F_9$. (a) All occurrences of $F_{k-4}$ and the three matching
intervals with prefix $\mathtt{vu}$. The interval at 19 is the target itself; $s=f_{k-1}+f_{k-4}=26$
is the start of the third phrase. (b) A copy from 6 adds a red edge opposite to the existing edge
$6\to19$. (c) The copy from 27 is valid. Blue and green edges are references of the first and second
phrases, respectively, directed from copied positions to their sources.}
\label{fig:fib-second-combined}
\end{figure}

After the second phrase, the current position is $s=f_{k-1}+f_{k-4}$, and the remaining suffix has
length $f_{k-3}$. We show that $f_{k-3}-2$ characters can be validly copied from position $f_{k-1}$.
This establishes the existence of a valid copy of this length; it does not claim that the source at
$f_{k-1}$ is necessarily selected by the tie-breaking rule. We first verify that the source and
target substrings match, and then show that the proposed copy creates no directed cycle. First
consider the strings themselves. Since $F_k=F_{k-1}F_{k-2}$ and $F_{k-2}=F_{k-4}F_{k-5}F_{k-4}$, the
suffix starting at $s=f_{k-1}+f_{k-4}$ begins with $F_{k-5}F_{k-4}$. On the other hand, $F_{k-3}$
starts at position $f_{k-1}$. Since $k$ and $k-3$ have opposite parity,
Lemma~\ref{lem:fib-known-properties}(\ref{item:fib-commutation}) gives \[
F_{k-5}F_{k-4}=C_{k-3}\mathtt{uv}, \qquad F_{k-3}=C_{k-3}\mathtt{vu}. \] Thus the first $f_{k-3}-2$
characters agree, and hence the desired copy is possible at the level of string equality.

We next verify that this copy is valid. Consider the fixed integer interval $I=[s,f_k)$, of length
$f_{k-3}$, and define $\phi:I\to I$ by
\[
\phi(x)=
\begin{cases}
x+f_{k-3}-f_{k-4},&s\le x<s+f_{k-4},\\
x-f_{k-4},&s+f_{k-4}\le x<f_k.
\end{cases}
\]
The proposed copy adds the edge $x\to x-f_{k-4}$ for every $x\in I\setminus\{f_k-2,f_k-1\}$. For
each such $x$, we show that the reference path next reaches $I$ at $\phi(x)$, by considering the
following two cases.

\emph{Case (1): $s\le x<s+f_{k-4}$.} The source $x-f_{k-4}$ belongs to $[f_{k-1},s)$. The second phrase copies $[f_{k-1}-2,s)$ from $[f_k-f_{k-4}-2,f_k)$, so its references increase the position by $f_{k-3}$. Following this existing edge gives
\[
x\longrightarrow x-f_{k-4}\longrightarrow x-f_{k-4}+f_{k-3}=\phi(x)\in I.
\]

\emph{Case (2): $s+f_{k-4}\le x<f_k$.} The source $x-f_{k-4}$ already belongs to $I$. Thus a single new edge reaches $\phi(x)=x-f_{k-4}$.

Consequently, until the reference path reaches one of the last two positions, its successive visits
to $I$ are $x,\phi(x),\phi^2(x),\ldots$. We have $\phi(x)=s+((x-s-f_{k-4})\bmod f_{k-3})$ and, by
Lemma~\ref{lem:fib-known-properties}(\ref{item:fib-coprime}), $\gcd(f_{k-3},f_{k-4})=1$. Applying
Lemma~\ref{lem:fib-modular-shift} with $J=I$, $d=s$, $A=f_{k-3}$, and $B=f_{k-4}$, iteration of
$\phi$ visits every position of $I$ before returning to its starting position. The actual reference
path stops upon reaching $f_k-2$ or $f_k-1$, since these positions are not copied and remain
explicit one-character phrases. Before this copy, the dependency graph is acyclic, so any new cycle
must contain a newly added edge starting in $I\setminus\{f_k-2,f_k-1\}$. Every such path reaches an
explicit phrase before returning to its starting vertex. Thus no path containing a newly added edge
can form a directed cycle, and no new cycle is created. The proposed copy of length $f_{k-3}-2$ is
therefore valid, and the third phrase has length at least $f_{k-3}-2$ under every tie-breaking rule.

It remains to show that it cannot be longer. In any valid BMS, following references from a position
eventually reaches an explicit phrase containing the same character. Hence every distinct character
occurring in the string requires at least one explicit phrase. The first two phrases are copy
phrases, and $F_k$ contains both $a$ and $b$. If the third phrase had length at least $f_{k-3}-1$,
at most one position would remain for explicit phrases, which is impossible. Therefore the third
phrase has length exactly $f_{k-3}-2$. The remaining two characters are $\mathtt{u}$ and
$\mathtt{v}$, which are distinct. Since the first three finalized phrases are copy phrases, both
must be stored explicitly. This proves the stated phrase strings and $\zRR^\tau(F_k)=5$ for every
tie-breaking rule $\tau$.
\end{proof}

\begin{theorem}[Logarithmic separation from LZ77]\label{thm:fib-log-separation}
For every tie-breaking rule $\tau$, the family of Fibonacci words $(F_k)_{k\ge8}$ satisfies, with
$n=|F_k|$,
\[
\frac{z(F_k)}{\zRR^\tau(F_k)}=\Theta(\log n).
\]
Consequently, $z(T)=O(\zRR^\tau(T))$ does not hold in general.
\end{theorem}
\begin{proof}
Combine the known bound $z(F_k)=\Theta(\log |F_k|)$~\cite{navarro2021ordered} with
$\zRR^\tau(F_k)=5$ from Theorem~\ref{thm:fib-forward-constant}.
\end{proof}

\section{When LZ77 Is Smaller than LZRR}
\label{sec:bit_rev}

We use a family based on bit reversal to obtain a logarithmic separation with $z\ll\zRR$.

Let $h\ge3$ and $m=2^h$. Let $\pi_h$ be the $h$-bit bit-reversal permutation: $\pi_h(i)$ is the
integer obtained by reversing the $h$-bit binary representation of $i\in\{0,\ldots,m-1\}$ (with
leading zeros). More generally, for $d\ge0$ and $0\le t<2^d$, we write $\pi_d(t)=\NUM(\BIN_d(t)^R)$,
with $\pi_0(0)=0$. Let $\Sigma_h=\{a_i,b_i,\$_i:0\le i<m\}$, where all symbols are distinct. Fix the
alphabet order $\$_{m-1}\prec \cdots\prec \$_1\prec \$_0\prec a_0\prec \cdots\prec a_{m-1}\prec
b_0\prec \cdots\prec b_{m-1}.$ For each $0\le x<m$, define a block $B_x$ of length $m+1$ by
\[
B_x[i]=\begin{cases}a_i,&\pi_h(i)<x,\\b_i,&\pi_h(i)\ge x\end{cases}
\quad(0\le i<m),\qquad B_x[m]=\$_x.
\]
We call $B_x[0..m-1]$ the \emph{body} of $B_x$. The position $B_x[i]$ has \emph{offset $i$} within
the body, with offsets numbered from zero.

For $x\in\{0,\ldots,m-1\}$, let $\LSB(x)$ be the number of trailing zeros in its $h$-bit binary
representation. In particular, $\LSB(0)=h$. For each $0\le k\le h$, list the indices with
$\LSB(x)=k$ in decreasing order:
\[
x_{k,1}>x_{k,2}>\cdots>x_{k,q_k},\qquad
q_k=\begin{cases}m/2^{k+1},&0\le k<h,\\1,&k=h.\end{cases}
\]
Define $H_k=B_{x_{k,1}}\cdots B_{x_{k,q_k}},  T_h=H_0H_1\cdots H_{h-1}H_h.$

That is, we concatenate the blocks in increasing order of $\LSB(x)$, and in decreasing order of $x$
among blocks with the same LSB value. We call $H_k$ level $k$. The last level is $H_h=B_0$. For
example, $T_3=B_7B_5B_3B_1B_6B_2B_4B_0.$ Let $t_x$ be the starting position of $B_x$ in $T_h$, so
$T_h[t_x..t_x+m]=B_x$. Each of the $m$ blocks occurs exactly once, so $n=|T_h|=m(m+1),
m=\Theta(\sqrt n),  h=\Theta(\log n).$ Our string family modifies the bit-reversal instance of
Shibata and Fujie~\cite{shibata2026lz}. Bille et al.~\cite{bille2026endvs77} also use a construction
that concatenates blocks defined by a bit-reversal permutation.

\begin{samepage}
\begin{theorem}[Logarithmic separation from LZ77]\label{thm:bitrev-separation}
For the string family above, the following bounds hold:
\begin{enumerate}
\item[(i)] $z(T_h)\le6m-5$.
\item[(ii)] Under the original tie-breaking rule (Definition~\ref{def:lzrr-candidate-order}),
$\zRR(T_h)\ge mh/2+1$.
\end{enumerate}
Hence $\zRR(T_h)=\Omega(\sqrt n\log n)$ and $\zRR(T_h)/z(T_h)=\Theta(\log n)$.
\end{theorem}
\end{samepage}

We first establish basic properties of the blocks, and then prove parts (i) and (ii) of
Theorem~\ref{thm:bitrev-separation}.

We use the following properties, which correspond to Properties 3 and 4 of~\cite{shibata2026lz}.
For each offset $i$, call $\pi_h(i)$ its \emph{threshold}: $B_x[i]=b_i$ for $x\le\pi_h(i)$ and
$B_x[i]=a_i$ for $x>\pi_h(i)$. Thus the character changes from $b_i$ to $a_i$ between block indices
$\pi_h(i)$ and $\pi_h(i)+1$, when both indices are in range.

\begin{figure}[htbp]
\centering
\begin{tikzpicture}[x=.82cm,y=.54cm,font=\small,>=Stealth]
\node[anchor=east] at (-.5,.8) {$\text{Offset } i$};
\node[anchor=east] at (-.5,0) {$\pi_3(i)$};

\node at (8.15,.8) {Delimiter};
\node at (0,.8) {$ 0 $}; \node at (0,0) {$ 0 $};
\node at (1,.8) {$ 1 $}; \node at (1,0) {$ 4 $};
\node at (2,.8) {$ 2 $}; \node at (2,0) {$ 2 $};
\node at (3,.8) {$ 3 $}; \node at (3,0) {$ 6 $};
\node at (4,.8) {$ 4 $}; \node at (4,0) {$ 1 $};
\node at (5,.8) {$ 5 $}; \node at (5,0) {$ 5 $};
\node at (6,.8) {$ 6 $}; \node at (6,0) {$ 3 $};
\node at (7,.8) {$ 7 $}; \node at (7,0) {$ 7 $};
\node[anchor=east] at (-.5,-1) {$B_0$};
\filldraw[fill=black!15,draw=white] (0-.48,-1-.47) rectangle (0+.48,-1+.47);\node at (0,-1) {$ b_0 $};
\filldraw[fill=black!15,draw=white] (1-.48,-1-.47) rectangle (1+.48,-1+.47);\node at (1,-1) {$ b_1 $};
\filldraw[fill=black!15,draw=white] (2-.48,-1-.47) rectangle (2+.48,-1+.47);\node at (2,-1) {$ b_2 $};
\filldraw[fill=black!15,draw=white] (3-.48,-1-.47) rectangle (3+.48,-1+.47);\node at (3,-1) {$ b_3 $};
\filldraw[fill=black!15,draw=white] (4-.48,-1-.47) rectangle (4+.48,-1+.47);\node at (4,-1) {$ b_4 $};
\filldraw[fill=black!15,draw=white] (5-.48,-1-.47) rectangle (5+.48,-1+.47);\node at (5,-1) {$ b_5 $};
\filldraw[fill=black!15,draw=white] (6-.48,-1-.47) rectangle (6+.48,-1+.47);\node at (6,-1) {$ b_6 $};
\filldraw[fill=black!15,draw=white] (7-.48,-1-.47) rectangle (7+.48,-1+.47);\node at (7,-1) {$ b_7 $};
\node at (8.15,-1) {$\$_0$};
\node[anchor=east] at (-.5,-2) {$B_1$};
\filldraw[fill=blue!12,draw=white] (0-.48,-2-.47) rectangle (0+.48,-2+.47);\node at (0,-2) {$ a_0 $};
\filldraw[fill=black!15,draw=white] (1-.48,-2-.47) rectangle (1+.48,-2+.47);\node at (1,-2) {$ b_1 $};
\filldraw[fill=black!15,draw=white] (2-.48,-2-.47) rectangle (2+.48,-2+.47);\node at (2,-2) {$ b_2 $};
\filldraw[fill=black!15,draw=white] (3-.48,-2-.47) rectangle (3+.48,-2+.47);\node at (3,-2) {$ b_3 $};
\filldraw[fill=black!15,draw=white] (4-.48,-2-.47) rectangle (4+.48,-2+.47);\node at (4,-2) {$ b_4 $};
\filldraw[fill=black!15,draw=white] (5-.48,-2-.47) rectangle (5+.48,-2+.47);\node at (5,-2) {$ b_5 $};
\filldraw[fill=black!15,draw=white] (6-.48,-2-.47) rectangle (6+.48,-2+.47);\node at (6,-2) {$ b_6 $};
\filldraw[fill=black!15,draw=white] (7-.48,-2-.47) rectangle (7+.48,-2+.47);\node at (7,-2) {$ b_7 $};
\node at (8.15,-2) {$\$_1$};
\node[anchor=east] at (-.5,-3) {$B_2$};
\filldraw[fill=blue!12,draw=white] (0-.48,-3-.47) rectangle (0+.48,-3+.47);\node at (0,-3) {$ a_0 $};
\filldraw[fill=black!15,draw=white] (1-.48,-3-.47) rectangle (1+.48,-3+.47);\node at (1,-3) {$ b_1 $};
\filldraw[fill=black!15,draw=white] (2-.48,-3-.47) rectangle (2+.48,-3+.47);\node at (2,-3) {$ b_2 $};
\filldraw[fill=black!15,draw=white] (3-.48,-3-.47) rectangle (3+.48,-3+.47);\node at (3,-3) {$ b_3 $};
\filldraw[fill=blue!12,draw=white] (4-.48,-3-.47) rectangle (4+.48,-3+.47);\node at (4,-3) {$ a_4 $};
\filldraw[fill=black!15,draw=white] (5-.48,-3-.47) rectangle (5+.48,-3+.47);\node at (5,-3) {$ b_5 $};
\filldraw[fill=black!15,draw=white] (6-.48,-3-.47) rectangle (6+.48,-3+.47);\node at (6,-3) {$ b_6 $};
\filldraw[fill=black!15,draw=white] (7-.48,-3-.47) rectangle (7+.48,-3+.47);\node at (7,-3) {$ b_7 $};
\node at (8.15,-3) {$\$_2$};
\node[anchor=east] at (-.5,-4) {$B_3$};
\filldraw[fill=blue!12,draw=white] (0-.48,-4-.47) rectangle (0+.48,-4+.47);\node at (0,-4) {$ a_0 $};
\filldraw[fill=black!15,draw=white] (1-.48,-4-.47) rectangle (1+.48,-4+.47);\node at (1,-4) {$ b_1 $};
\filldraw[fill=blue!12,draw=white] (2-.48,-4-.47) rectangle (2+.48,-4+.47);\node at (2,-4) {$ a_2 $};
\filldraw[fill=black!15,draw=white] (3-.48,-4-.47) rectangle (3+.48,-4+.47);\node at (3,-4) {$ b_3 $};
\filldraw[fill=blue!12,draw=white] (4-.48,-4-.47) rectangle (4+.48,-4+.47);\node at (4,-4) {$ a_4 $};
\filldraw[fill=black!15,draw=white] (5-.48,-4-.47) rectangle (5+.48,-4+.47);\node at (5,-4) {$ b_5 $};
\filldraw[fill=black!15,draw=white] (6-.48,-4-.47) rectangle (6+.48,-4+.47);\node at (6,-4) {$ b_6 $};
\filldraw[fill=black!15,draw=white] (7-.48,-4-.47) rectangle (7+.48,-4+.47);\node at (7,-4) {$ b_7 $};
\node at (8.15,-4) {$\$_3$};
\node[anchor=east] at (-.5,-5) {$B_4$};
\filldraw[fill=blue!12,draw=white] (0-.48,-5-.47) rectangle (0+.48,-5+.47);\node at (0,-5) {$ a_0 $};
\filldraw[fill=black!15,draw=white] (1-.48,-5-.47) rectangle (1+.48,-5+.47);\node at (1,-5) {$ b_1 $};
\filldraw[fill=blue!12,draw=white] (2-.48,-5-.47) rectangle (2+.48,-5+.47);\node at (2,-5) {$ a_2 $};
\filldraw[fill=black!15,draw=white] (3-.48,-5-.47) rectangle (3+.48,-5+.47);\node at (3,-5) {$ b_3 $};
\filldraw[fill=blue!12,draw=white] (4-.48,-5-.47) rectangle (4+.48,-5+.47);\node at (4,-5) {$ a_4 $};
\filldraw[fill=black!15,draw=white] (5-.48,-5-.47) rectangle (5+.48,-5+.47);\node at (5,-5) {$ b_5 $};
\filldraw[fill=blue!12,draw=white] (6-.48,-5-.47) rectangle (6+.48,-5+.47);\node at (6,-5) {$ a_6 $};
\filldraw[fill=black!15,draw=white] (7-.48,-5-.47) rectangle (7+.48,-5+.47);\node at (7,-5) {$ b_7 $};
\node at (8.15,-5) {$\$_4$};
\node[anchor=east] at (-.5,-6) {$B_5$};
\filldraw[fill=blue!12,draw=white] (0-.48,-6-.47) rectangle (0+.48,-6+.47);\node at (0,-6) {$ a_0 $};
\filldraw[fill=blue!12,draw=white] (1-.48,-6-.47) rectangle (1+.48,-6+.47);\node at (1,-6) {$ a_1 $};
\filldraw[fill=blue!12,draw=white] (2-.48,-6-.47) rectangle (2+.48,-6+.47);\node at (2,-6) {$ a_2 $};
\filldraw[fill=black!15,draw=white] (3-.48,-6-.47) rectangle (3+.48,-6+.47);\node at (3,-6) {$ b_3 $};
\filldraw[fill=blue!12,draw=white] (4-.48,-6-.47) rectangle (4+.48,-6+.47);\node at (4,-6) {$ a_4 $};
\filldraw[fill=black!15,draw=white] (5-.48,-6-.47) rectangle (5+.48,-6+.47);\node at (5,-6) {$ b_5 $};
\filldraw[fill=blue!12,draw=white] (6-.48,-6-.47) rectangle (6+.48,-6+.47);\node at (6,-6) {$ a_6 $};
\filldraw[fill=black!15,draw=white] (7-.48,-6-.47) rectangle (7+.48,-6+.47);\node at (7,-6) {$ b_7 $};
\node at (8.15,-6) {$\$_5$};
\node[anchor=east] at (-.5,-7) {$B_6$};
\filldraw[fill=blue!12,draw=white] (0-.48,-7-.47) rectangle (0+.48,-7+.47);\node at (0,-7) {$ a_0 $};
\filldraw[fill=blue!12,draw=white] (1-.48,-7-.47) rectangle (1+.48,-7+.47);\node at (1,-7) {$ a_1 $};
\filldraw[fill=blue!12,draw=white] (2-.48,-7-.47) rectangle (2+.48,-7+.47);\node at (2,-7) {$ a_2 $};
\filldraw[fill=black!15,draw=white] (3-.48,-7-.47) rectangle (3+.48,-7+.47);\node at (3,-7) {$ b_3 $};
\filldraw[fill=blue!12,draw=white] (4-.48,-7-.47) rectangle (4+.48,-7+.47);\node at (4,-7) {$ a_4 $};
\filldraw[fill=blue!12,draw=white] (5-.48,-7-.47) rectangle (5+.48,-7+.47);\node at (5,-7) {$ a_5 $};
\filldraw[fill=blue!12,draw=white] (6-.48,-7-.47) rectangle (6+.48,-7+.47);\node at (6,-7) {$ a_6 $};
\filldraw[fill=black!15,draw=white] (7-.48,-7-.47) rectangle (7+.48,-7+.47);\node at (7,-7) {$ b_7 $};
\node at (8.15,-7) {$\$_6$};
\node[anchor=east] at (-.5,-8) {$B_7$};
\filldraw[fill=blue!12,draw=white] (0-.48,-8-.47) rectangle (0+.48,-8+.47);\node at (0,-8) {$ a_0 $};
\filldraw[fill=blue!12,draw=white] (1-.48,-8-.47) rectangle (1+.48,-8+.47);\node at (1,-8) {$ a_1 $};
\filldraw[fill=blue!12,draw=white] (2-.48,-8-.47) rectangle (2+.48,-8+.47);\node at (2,-8) {$ a_2 $};
\filldraw[fill=blue!12,draw=white] (3-.48,-8-.47) rectangle (3+.48,-8+.47);\node at (3,-8) {$ a_3 $};
\filldraw[fill=blue!12,draw=white] (4-.48,-8-.47) rectangle (4+.48,-8+.47);\node at (4,-8) {$ a_4 $};
\filldraw[fill=blue!12,draw=white] (5-.48,-8-.47) rectangle (5+.48,-8+.47);\node at (5,-8) {$ a_5 $};
\filldraw[fill=blue!12,draw=white] (6-.48,-8-.47) rectangle (6+.48,-8+.47);\node at (6,-8) {$ a_6 $};
\filldraw[fill=black!15,draw=white] (7-.48,-8-.47) rectangle (7+.48,-8+.47);\node at (7,-8) {$ b_7 $};
\node at (8.15,-8) {$\$_7$};
\draw[blue!75!black,very thick] (1.5,-.5) rectangle (2.5,-8.5);
\draw[blue!75!black,very thick] (1.5,-3.5)--(2.5,-3.5);
\end{tikzpicture}
\caption{Blocks and thresholds for $h=3$. Each row is a block $B_x$ in increasing index order, and
each column $i$ is an offset counted from zero at the start of the block. The entry $B_x[i]$ is the
character at position $t_x+i$ in $T_3$. At offset $i$, the character is $b_i$ if $x\le\pi_3(i)$ and
$a_i$ if $x>\pi_3(i)$. The blue outline marks offset 2, where the character changes between $B_2$
and $B_3$. The table layout follows~\cite[Fig.~1]{shibata2026lz}
and~\cite[Fig.~3]{bille2026endvs77}.}
\label{fig:bitrev-offset-threshold}
\end{figure}

\begin{property}\label{prop:bitrev-alignment}
Let $0\le l_x\le r_x<m$ and $0\le l_y\le r_y<m$. Matching substrings within bodies have the same
offset interval; namely, if $B_x[l_x..r_x]=B_y[l_y..r_y]$, then $(l_x,r_x)=(l_y,r_y)$. If these are
occurrences at distinct positions in $T_h$, then $x\ne y$.
\end{property}
\begin{proof}
The symbols $a_i$ and $b_i$ occur only at offset $i$. Equality of the first characters therefore
gives $l_x=l_y$. Equality of the substring lengths then gives $r_x=r_y$. Occurrences at distinct
positions therefore belong to distinct blocks.
\end{proof}
\begin{property}\label{prop:bitrev-mismatch}
For $0\le x<y<m$ and $0\le i<m$, we have $B_x[i]\ne B_y[i]$ if and only if $x\le\pi_h(i)<y$.
\end{property}
\begin{proof}
By definition, a mismatch occurs precisely when $B_x[i]=b_i$ and $B_y[i]=a_i$. This is equivalent to
$x\le\pi_h(i)<y$.
\end{proof}
\begin{property}\label{prop:bitrev-body-distance}
For any $0\le x,y<m$, the Hamming distance between the bodies of $B_x$ and $B_y$, namely the number
of offsets with different characters, is $|x-y|$.
\end{property}
\begin{proof}
For $x<y$, Property~\ref{prop:bitrev-mismatch} and the fact that $\pi_h$ is a permutation give
exactly $y-x$ mismatching offsets. The case $x>y$ is symmetric, and the distance is zero for $x=y$.
\end{proof}
A substring containing a unique delimiter cannot occur twice. In any valid BMS of $T_h$ (in
particular, in LZ77 and LZRR), every copy phrase is contained in one body, and its source starts at
the same offset by Property~\ref{prop:bitrev-alignment}. Every delimiter is an explicit
one-character phrase.

\begin{lemma}[Upper bound for LZ77]\label{lem:bitrev-lz-upper}
$z(T_h)\le6m-5$.
\end{lemma}
\begin{proof}
The first block is $B_{m-1}$ and can be parsed into $m+1$ explicit phrases. Consider a remaining
block $B_x$, where $0\le x\le m-2$. If $x$ is even, then $x+1$ is odd, so $B_{x+1}$ is in level 0
and occurs before $B_x$ in $T_h$. By Property~\ref{prop:bitrev-body-distance}, their bodies differ
in exactly one offset. We can parse $B_x$ into at most four phrases: at most two copy phrases
referring to the matching parts of $B_{x+1}$, one explicit phrase for the mismatching character, and
one for the delimiter.

If $x$ is odd, then $x+2$ is odd as well. Both blocks belong to level 0, where blocks are
concatenated in decreasing index order. Since $x+2>x$, the block $B_{x+2}$ occurs before $B_x$. By
Property~\ref{prop:bitrev-body-distance}, their bodies differ in exactly two offsets. We can
therefore parse $B_x$ into at most six phrases: at most three copy phrases referring to the matching
parts of $B_{x+2}$, two explicit phrases for the mismatching characters, and one for the delimiter.

Every copy phrase in this parsing refers to the left. The greedy LZ77 parsing minimizes the phrase
count among such parsings~\cite{navarro2021ordered}. There are $m/2$ even and $m/2-1$ odd indices in
$\{0,\ldots,m-2\}$, so $z(T_h)\le(m+1)+4\left(\frac m2\right)+6\left(\frac
m2-1\right)=6m-5.\qedhere$
\end{proof}

Let $1\le x<m$ and $k=\LSB(x)<h$. For the level-$k$ block $B_x$ and an integer
$s\in\{0,\ldots,2^k-1\}$, define
\begin{equation}
l_{x,s}=s\,2^{h-k}+v_x,\qquad r_{x,s}=l_{x,s}+2^{h-k-1},
\label{eq:bitrev-intervals}
\end{equation}
where $v_x=\pi_{h-k-1}\!\left(\left\lfloor\frac{x}{2^{k+1}}\right\rfloor\right).$

Since $x$ ends in $10^k$ when written in $h$ bits, $v_x$ is obtained by removing these last $k+1$
bits and reversing the remaining $h-k-1$ bits.

Consider the interval $I_{x,s}=[l_{x,s},r_{x,s}]$, and write $B_x[I_{x,s}]=B_x[l_{x,s}..r_{x,s}]$
for the corresponding substring. Since $0\le v_x<2^{h-k-1}$, these intervals are contained in the
body. For $0\le s<2^k-1$, $l_{x,s+1}-l_{x,s}=2^{h-k}$, while $r_{x,s}-l_{x,s}=2^{h-k-1}$ for every
$s$. Hence the $2^k$ intervals in $B_x$ are pairwise disjoint. There are $m/2^{k+1}$ blocks in level
$k$, so the total number of intervals is
\[
\sum_{k=0}^{h-1}\frac{m}{2^{k+1}}\,2^k=\frac{mh}{2}.
\]
To prove Theorem~\ref{thm:bitrev-separation}(ii), we show that no LZRR phrase contains the whole
text interval $[t_x+l_{x,s},t_x+r_{x,s}]$. We first determine the block indices $y\ne x$ for which
$B_y[I_{x,s}]=B_x[I_{x,s}]$. We then prove that copying these offsets from any such block creates a
directed cycle in the dependency graph.

For convenience, write $L_{x,s}=\pi_h(l_{x,s})$ and $R_{x,s}=\pi_h(r_{x,s})$ for the thresholds of
the left and right endpoint offsets of $I_{x,s}$. These thresholds characterize exactly which block
indices $y$ satisfy $B_y[I_{x,s}]=B_x[I_{x,s}]$.

\begin{lemma}\label{lem:bitrev-window}
For every integer $0\le y<m$,
\begin{equation}
B_y[l_{x,s}..r_{x,s}]=B_x[l_{x,s}..r_{x,s}]
\quad\Longleftrightarrow\quad L_{x,s}<y\le R_{x,s}.
\label{eq:bitrev-window}
\end{equation}
\end{lemma}
\begin{proof}
By definition, $\BIN_{h-k-1}(v_x)^R=\BIN_{h-k-1}(\lfloor x/2^{k+1}\rfloor)$. For binary strings
$u,v$, we have $(uv)^R=v^Ru^R$ and $\NUM(uv)=2^{|v|}\NUM(u)+\NUM(v)$. Since $0\le v_x<2^{h-k-1}$,
the lowest $h-k$ bits of $l_{x,s}$ are $0\BIN_{h-k-1}(v_x)$. Adding $2^{h-k-1}$ to obtain $r_{x,s}$
changes only this leading $0$ to $1$. Thus,
\[
\begin{aligned}
\BIN_h(l_{x,s})&=\BIN_k(s)\,0\,\BIN_{h-k-1}(v_x),\\
\BIN_h(r_{x,s})&=\BIN_k(s)\,1\,\BIN_{h-k-1}(v_x).
\end{aligned}
\]
\[
\begin{aligned}
L_{x,s}&=\NUM\bigl(\BIN_h(l_{x,s})^R\bigr)&&\text{\footnotesize (definition of $\pi_h$)}\\
&=\NUM\bigl((\BIN_k(s)\,0\,\BIN_{h-k-1}(v_x))^R\bigr)&&\text{\footnotesize (binary form of $l_{x,s}$)}\\
&=2^{k+1}\NUM\bigl(\BIN_{h-k-1}(v_x)^R\bigr)+\pi_k(s)&&\text{\footnotesize (reversal and place values)}\\
&=2^{k+1}\left\lfloor\frac{x}{2^{k+1}}\right\rfloor+\pi_k(s)&&\text{\footnotesize (definition of $v_x$)}\\
&=x-2^k+\pi_k(s)&&\text{\footnotesize ($x$ ends in $10^k$)}\\
R_{x,s}&=\NUM\bigl(\BIN_h(r_{x,s})^R\bigr)&&\text{\footnotesize (definition of $\pi_h$)}\\
&=\NUM\bigl((\BIN_k(s)\,1\,\BIN_{h-k-1}(v_x))^R\bigr)&&\text{\footnotesize (binary form of $r_{x,s}$)}\\
&=L_{x,s}+2^k&&\text{\footnotesize (the changed bit has weight $2^k$)}\\
&=x+\pi_k(s)&&\text{\footnotesize (value of $L_{x,s}$)}
\end{aligned}
\]
The last simplification uses $x=2^{k+1}\lfloor x/2^{k+1}\rfloor+2^k$, since $\LSB(x)=k$. Hence,
\begin{equation}
L_{x,s}=x-2^k+\pi_k(s),\qquad R_{x,s}=x+\pi_k(s).
\label{eq:bitrev-endpoint-thresholds}
\end{equation}
In particular, $L_{x,s}<x\le R_{x,s}$, so $B_x[l_{x,s}]=a_{l_{x,s}}$ and $B_x[r_{x,s}]=b_{r_{x,s}}$.
If $B_x[I_{x,s}]=B_y[I_{x,s}]$, both endpoint characters must match. By the definition of $B_y$,
this implies $L_{x,s}<y\le R_{x,s}$.

Conversely, suppose that $L_{x,s}<y\le R_{x,s}$. The endpoint characters match by the argument
above. For any integer $l_{x,s}<i<r_{x,s}$,
\[
\begin{aligned}
\BIN_h(i)&=\BIN_k(s)\,\BIN_{h-k}\bigl(i-s2^{h-k}\bigr)&&\text{\footnotesize (bit decomposition)}\\
\pi_h(i)&=2^k\NUM\!\left(\BIN_{h-k}\bigl(i-s2^{h-k}\bigr)^R\right)+\pi_k(s)&&\text{\footnotesize (reversal and place values)}
\end{aligned}
\]
These identities and the definitions of $L_{x,s},R_{x,s}$ give $\pi_h(i)\equiv L_{x,s}\equiv
R_{x,s}\equiv\pi_k(s)\pmod{2^k}$. Injectivity of $\pi_h$ and $i\ne l_{x,s}$ imply $\pi_h(i)\ne
L_{x,s}$. The congruence above gives $|\pi_h(i)-L_{x,s}|\ge2^k$. Together with
$R_{x,s}=L_{x,s}+2^k$, this yields $\pi_h(i)<L_{x,s}$ or $R_{x,s}\le\pi_h(i)$. In the first case,
$B_x[i]=B_y[i]=a_i$; in the second, $B_x[i]=B_y[i]=b_i$. All interior characters also match. This
proves the lemma.
\end{proof}

\begin{lemma}\label{lem:bitrev-order}
If $L_{x,s}<y\le R_{x,s}$ and $y\ne x$, then $B_y$ occurs to the left of $B_x$ in $T_h$. In other
words, $B_x$ is the rightmost block in this range.
\end{lemma}
\begin{proof}
By~\eqref{eq:bitrev-endpoint-thresholds}, the integers in this range are $x-2^k+\pi_k(s)+1,\
\ldots,\ x+\pi_k(s).$ There are $2^k$ of them. Since $0\le\pi_k(s)<2^k$, the range contains $x$.
Among $2^k$ consecutive integers there is exactly one multiple of $2^k$, which here is $x$.
Therefore $y\ne x$ implies $\LSB(y)<k$. Since levels are concatenated in increasing order of LSB
value, $B_y$ occurs before $B_x$.
\end{proof}

\begin{figure}[!htbp]
\centering
\includegraphics[width=\linewidth]{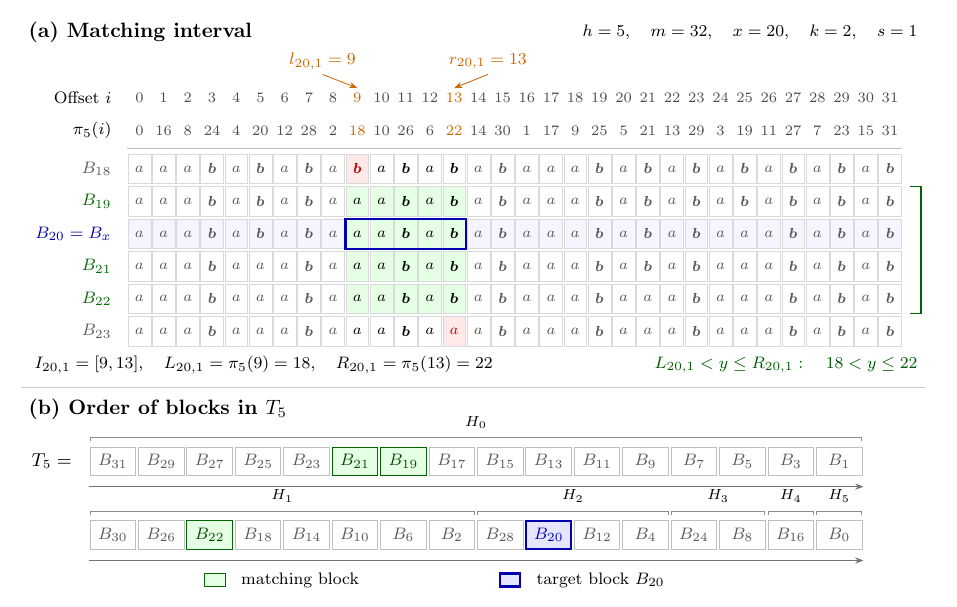}
\caption{Illustration of Lemmas~\ref{lem:bitrev-window} and~\ref{lem:bitrev-order} for $h=5$,
$x=20$, and $s=1$.
(a) The interval $I_{20,1}=[9,13]$ has endpoint thresholds $L_{20,1}=18$ and $R_{20,1}=22$.
The substring $B_y[9..13]$ matches $B_{20}[9..13]$ exactly when $18<y\le22$.
Character subscripts are omitted, with $b$ shown in bold.
(b) The blocks of $T_5$ are shown in order, from left to right along the upper row and then the
lower row.
The other blocks containing a matching interval, $B_{21}$, $B_{19}$, and $B_{22}$, all precede
$B_{20}$.
Green marks matching intervals and their blocks, blue outlines the target, and red marks mismatching
endpoints.}
\label{fig:bitrev-window-order-example}
\end{figure}

Figure~\ref{fig:bitrev-window-order-example} illustrates the matching window and block order in
Lemmas~\ref{lem:bitrev-window} and~\ref{lem:bitrev-order}.

Together, Lemmas~\ref{lem:bitrev-window} and~\ref{lem:bitrev-order} show that every block $B_y$
other than $B_x$ that can be a source for all of $B_x[I_{x,s}]$ satisfies $L_{x,s}<y\le R_{x,s}$ and
has already been parsed when parsing of $B_x$ begins. We next show that copying from a block in this
range creates a directed cycle in the dependency graph. We first establish properties of suffix
order and LCPs that determine the candidate order in Definition~\ref{def:lzrr-candidate-order}.

\begin{lemma}\label{lem:bitrev-monotone}
If $0\le i<m$, $0\le y<z<m$, and $B_y[i]=B_z[i]$, then
\begin{equation}
T_h[t_z+i..]\prec T_h[t_y+i..].
\label{eq:bitrev-suffix-order}
\end{equation}
Moreover, for distinct $y,z$ and any $x$ strictly between them,
\begin{equation}
\lcp_{T_h}(t_y+i,t_x+i)\ge\lcp_{T_h}(t_y+i,t_z+i).
\label{eq:bitrev-lcp-monotonicity}
\end{equation}
\end{lemma}
\begin{proof}
We first prove~\eqref{eq:bitrev-suffix-order}. The first mismatch between $T_h[t_z+i..]$ and
$T_h[t_y+i..]$ is either at the delimiter or at a offset $j$ with $i\le j<m$. If the first mismatch
is at the delimiter, $\$_z\prec\$_y$ gives~\eqref{eq:bitrev-suffix-order}. Otherwise,
Property~\ref{prop:bitrev-mismatch} gives $y\le\pi_h(j)<z$, so $B_z[j]=a_j\prec b_j=B_y[j]$. This
also proves~\eqref{eq:bitrev-suffix-order}.

We next prove~\eqref{eq:bitrev-lcp-monotonicity}. Let $\ell=\lcp_{T_h}(t_y+i,t_z+i)$. Since
$\$_y\ne\$_z$, every offset $j$ with $i\le j<i+\ell$ satisfies $j<m$ and $B_y[j]=B_z[j]$. Assume
$y<x<z$; the other case is symmetric. If both characters are $a_j$, then $\pi_h(j)<y<x$. If both are
$b_j$, then $\pi_h(j)\ge z>x$. In either case, $B_x[j]=B_y[j]=B_z[j]$. The suffixes at $t_y+i$ and
$t_x+i$ share a prefix of length $\ell$. This proves~\eqref{eq:bitrev-lcp-monotonicity}.
\end{proof}

\begin{lemma}\label{lem:bitrev-no-jump}
Suppose that an LZRR copy phrase $T_h[t_y+i..t_y+i+\ell-1]$ of length $\ell$ starts at $t_y+i$ and
refers to $t_z+i$. Then no block $B_x$ with $t_x>t_y$ has an index strictly between $y$ and $z$.
\end{lemma}
\begin{proof}
Suppose that there is an $x$ with $t_x>t_y$ and either $y<x<z$ or $z<x<y$. Let $\mathcal P$ be the
parsing just before the algorithm determines the source and length of this phrase. Since
$t_x+i>t_y+i$, Lemmas~\ref{lem:future-copy} and~\ref{lem:bitrev-monotone} give
\[
\begin{aligned}
\LTP(\mathcal P,t_y+i,t_x+i)
 &=\lcp_{T_h}(t_y+i,t_x+i)\\
 &\ge\lcp_{T_h}(t_y+i,t_z+i)
 \ge\LTP(\mathcal P,t_y+i,t_z+i).
\end{aligned}
\]
If the LCP with $t_x+i$ is strictly longer, the candidate $t_x+i$ is examined before $t_z+i$ by
Definition~\ref{def:lzrr-candidate-order}. If the LCPs are equal, the rank $\ISA[t_x+i]$ is strictly
between $\ISA[t_y+i]$ and $\ISA[t_z+i]$ by Lemma~\ref{lem:bitrev-monotone}. Thus both candidate
ranks are on the same side of $\ISA[t_y+i]$ in the suffix array. The definition of $\dist$ subtracts
0 from both rank distances if the candidates are to the right of this rank, and $1/2$ from both if
they are to the left. This subtraction does not change their comparison. We obtain
$\dist_{t_y+i}(t_x+i)<\dist_{t_y+i}(t_z+i)$, so $t_x+i$ is examined first in this case as well.

If the search stops before reaching $t_z+i$, that candidate is not selected. Otherwise, the best
copy length already found is at least
\[
\LTP(\mathcal P,t_y+i,t_x+i)\ge\LTP(\mathcal P,t_y+i,t_z+i).
\]
The candidate $t_z+i$ cannot strictly improve this value, so it cannot be selected by the update
rule. This contradicts the assumed source of the phrase.
\end{proof}

\begin{lemma}\label{lem:bitrev-components}
Let $1\le x<m$ and $k=\LSB(x)$. Let $\mathcal P_x$ be the parsing immediately before LZRR starts
processing $B_x$, and let $G_{T_h,t_x}=G_{\mathcal P_x}$ be its dependency graph. For every integer
$0\le s<2^k$,
\begin{equation}
t_y+l_{x,s}\Dep_{\mathcal P_x}t_x+l_{x,s}\qquad(L_{x,s}<y\le x),
\label{eq:bitrev-left-component}
\end{equation}
\begin{equation}
t_y+r_{x,s}\Dep_{\mathcal P_x}t_x+r_{x,s}\qquad(x\le y\le R_{x,s}).
\label{eq:bitrev-right-component}
\end{equation}
\end{lemma}
\begin{proof}
Fix $s$. We first prove~\eqref{eq:bitrev-left-component}. Consider the vertex set $S_{x,s}^{\mathrm
L}=\{t_y+l_{x,s}:L_{x,s}<y\le x\}$. The position $t_x+l_{x,s}$ is still an explicit one-character
phrase of $\mathcal P_x$ and has outdegree zero in $G_{T_h,t_x}$. For $y\ne x$, the block $B_y$
occurs to the left of $B_x$ by Lemma~\ref{lem:bitrev-order}. The phrase containing $t_y+l_{x,s}$ has
already been processed, and its source and length have been determined. It must be a copy phrase.
Otherwise, when LZRR processed this phrase, the candidate $t_x+l_{x,s}$ to its right would have
given a copy of length at least one by Lemma~\ref{lem:future-copy}, so LZRR would not have stored
the character explicitly. This vertex therefore has an outgoing edge.

Let $t_y+j$ be the start of its phrase. By Property~\ref{prop:bitrev-alignment}, the source starts
at some $t_z+j$, so the outgoing edge from $t_y+l_{x,s}$ leads to $t_z+l_{x,s}$. Since
$T_h[t_y+l_{x,s}]=a_{l_{x,s}}$, we have $B_z[l_{x,s}]=a_{l_{x,s}}$, and hence $z>L_{x,s}$. If $z>x$,
then $y<x<z$ and $t_x>t_y$. These inequalities contradict Lemma~\ref{lem:bitrev-no-jump} for the
phrase starting at $t_y+j$. We obtain $L_{x,s}<z\le x$, so the source also belongs to
$S_{x,s}^{\mathrm L}$.
All outgoing edges from the set remain within the set. In the subgraph induced by $S_{x,s}^{\mathrm
L}$, only $t_x+l_{x,s}$ has outdegree zero, and every other vertex has outdegree one. Every vertex
reaches $t_x+l_{x,s}$ by Lemma~\ref{lem:unique-sink-tree}. This
proves~\eqref{eq:bitrev-left-component}.

For the right endpoints, use $S_{x,s}^{\mathrm R}=\{t_y+r_{x,s}:x\le y\le R_{x,s}\}$. Vertex
$t_x+r_{x,s}$ has outdegree zero, and every other vertex has an outgoing edge for the same reason as
on the left. Write its source as $t_z+r_{x,s}$. Since $T_h[t_y+r_{x,s}]=b_{r_{x,s}}$, we have
$B_z[r_{x,s}]=b_{r_{x,s}}$, and hence $z\le R_{x,s}$. The case $z<x$ would contradict
Lemma~\ref{lem:bitrev-no-jump} for this phrase, so we obtain $x\le z\le R_{x,s}$. Thus the source
belongs to $S_{x,s}^{\mathrm R}$. Every vertex in this induced subgraph reaches $t_x+r_{x,s}$ by
Lemma~\ref{lem:unique-sink-tree}. This proves~\eqref{eq:bitrev-right-component}.
\end{proof}

\begin{figure}[!htbp]
\centering
\includegraphics[width=\linewidth]{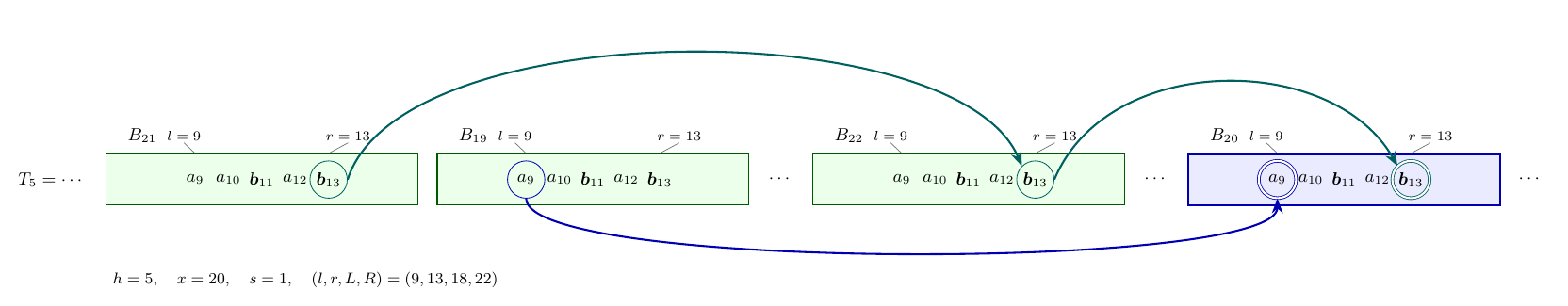}
\caption{Example of Lemma~\ref{lem:bitrev-components} for $h=5$, $x=20$, and $s=1$, using the same
blocks as Figure~\ref{fig:bitrev-window-order-example}.
The boxes are arranged in their order in $T_5$. Only the characters at offsets $9$ through $13$ are
displayed; blank margins represent the omitted prefix and suffix of each block.
Arrows are existing dependency edges under the original tie-breaking rule, immediately before LZRR
processes $B_{20}$.
The left endpoints with $18<y\le20$ and the right endpoints with $20\le y\le22$ reach the
corresponding endpoints of $B_{20}$, marked by double circles.
These unprocessed vertices have no outgoing edges. Other edges are omitted.}
\label{fig:bitrev-endpoint-paths}
\end{figure}

Figure~\ref{fig:bitrev-endpoint-paths} illustrates the two reachability relations in
Lemma~\ref{lem:bitrev-components}.

\begin{proof}[Proof of Theorem~\ref{thm:bitrev-separation}]
Part (i) follows from Lemma~\ref{lem:bitrev-lz-upper}. To prove (ii), suppose that an LZRR phrase
contains the whole text interval $[t_x+l_{x,s},t_x+r_{x,s}]$. Since the interval contains at least
two characters, this is a copy phrase. By Property~\ref{prop:bitrev-alignment} and
Lemma~\ref{lem:bitrev-window}, the source of this text interval occupies the same offsets in another
block $B_y$ with $L_{x,s}<y\le R_{x,s}$. If $y<x$, Lemma~\ref{lem:bitrev-components} gives
$t_y+l_{x,s}\Dep t_x+l_{x,s}$. If $y>x$, it gives $t_y+r_{x,s}\Dep t_x+r_{x,s}$. These relations
hold when the algorithm starts processing $B_x$ and are preserved when edges are added later.

When the algorithm starts processing the phrase that contains this text interval, both endpoints
$t_x+l_{x,s}$ and $t_x+r_{x,s}$ are still explicit one-character phrases and are the roots of their
components. If $y<x$, there is an existing directed path from the source $t_y+l_{x,s}$ to the root
$t_x+l_{x,s}$. Adding the edge $t_x+l_{x,s}\to t_y+l_{x,s}$ creates a directed cycle. If $y>x$, the
existing directed path from $t_y+r_{x,s}$ to $t_x+r_{x,s}$ and the new edge $t_x+r_{x,s}\to
t_y+r_{x,s}$ create a directed cycle. Thus the phrase is invalid, a contradiction. Consequently, no
LZRR phrase contains the whole text interval $[t_x+l_{x,s},t_x+r_{x,s}]$. Equivalently, for every
such interval, at least one LZRR phrase starts at a text position $p$ satisfying $t_x+l_{x,s}<p\le
t_x+r_{x,s}$.

These text intervals are pairwise disjoint, and there are $mh/2$ of them. They therefore contain at
least $mh/2$ distinct phrase starts satisfying the inequalities above. Since the initial phrase
starts at position zero and is not among them, there are at least $mh/2+1$ phrases. This proves
(ii). Combining (i) and (ii), we obtain $\frac{\zRR(T_h)}{z(T_h)}\ge\frac{mh/2+1}{6m-5}>\frac
h{12}.$ Since $n=m(m+1)$, we have $m=\Theta(\sqrt n)$ and $h=\Theta(\log n)$, which give the claimed
asymptotic bounds. The general upper bound in Lemma~\ref{lem:z-zrr-general-ratios} gives
$\zRR(T_h)/z(T_h)=\Theta(\log n)$.
\end{proof}

\section{Conclusion}
\label{sec:conclusion}

In this paper, we showed that the numbers of phrases in LZ77 and LZRR can differ by a logarithmic
factor in either direction when both parsings are applied to the same orientation of the input. On
Fibonacci words, $z/\zRR^\tau=\Theta(\log n)$ for every tie-breaking rule $\tau$, while on our
bit-reversal-based family, $\zRR/z=\Theta(\log n)$ under the original rule of Nishimoto and Tabei.
Together with the general $O(\log n)$ upper bounds, these results determine the worst-case
separation between the two parsing sizes up to constant factors.

These worst-case separations do not contradict the empirical effectiveness of LZRR reported by
Nishimoto and Tabei, whose experiments showed that LZRR often produces fewer phrases than LZ77 on
benchmark collections~\cite{nishimoto2022lzrr}. Our results show that such an advantage is not
guaranteed asymptotically on every fixed orientation of the input.

Future work includes understanding the effect of tie-breaking on the LZRR parsing size, obtaining a
separation with $z\ll\zRR$ that is independent of tie-breaking, and reducing the alphabet size of
our construction. It is also natural to study the sensitivity of LZRR and its relation to other
repetitiveness measures and parsings that induce bidirectional macro schemes, such as $r$, $\chi$,
and $v$. On the algorithmic side, improving the $O(n^2\alpha(n))$-time algorithm for computing
LZRR~\cite{nishimoto2022lzrr}, and more generally designing faster algorithms for constructing small
bidirectional macro schemes, remain interesting problems.

\section*{AI Disclosure}
The research questions and candidate lower-bound constructions were proposed by the author.
OpenAI's ChatGPT and Anthropic's Claude were used to develop proofs of the main results
and to assist with writing, revising, and formatting the manuscript.
ChatGPT was also used to assist with creating and revising figures.
The author carefully verified all mathematical claims and proofs and takes full responsibility for
the correctness and originality of the paper.

\bibliographystyle{plainurl}
\bibliography{references}
\end{document}